\documentclass[aps,prb,reprint,amsfonts,amsmath,amssymb]{revtex4-1} % for arXiv
\usepackage[colorlinks=true,citecolor=blue,linkcolor=blue,filecolor=blue,urlcolor=blue]{hyperref} % hyperlink with color (blue)
\usepackage{bm} % bold
\usepackage{braket} % braket
\usepackage[pdftex]{graphicx} % figure for arXiv
\usepackage{multirow} % table
\usepackage{ulem} % line

\begin{document}

%%%%%%%%%%%%%%%%%%%%%%%%%%%%%%%%%%%%%%%%%%%%%%%%%%%%%%%%%%%%
%%%%%%%%%%%%%%% title and author information %%%%%%%%%%%%%%%
%%%%%%%%%%%%%%%%%%%%%%%%%%%%%%%%%%%%%%%%%%%%%%%%%%%%%%%%%%%%
% title
\title{Comparative study of the density matrix embedding theory for the Hubbard models}
% author information
\author{Masataka Kawano}
\email{kawano@g.ecc.u-tokyo.ac.jp}
\author{Chisa Hotta}
\affiliation{Department of Basic Science, University of Tokyo, Meguro-ku, Tokyo 153-8902, Japan}
\date{\today}

%%%%%%%%%%%%%%%%%%%%%%%%%%%%%%%%%%%%%%%%
%%%%%%%%%%%%%%% abstract %%%%%%%%%%%%%%%
%%%%%%%%%%%%%%%%%%%%%%%%%%%%%%%%%%%%%%%%
\begin{abstract}
We examine the performance of the density matrix embedding theory (DMET) recently proposed in
[G. Knizia and G. K.-L. Chan, \href{https://journals.aps.org/prl/abstract/10.1103/PhysRevLett.109.186404}{Phys. Rev. Lett. \textbf{109}, 186404 (2012)}].
The core of this method is to find a proper one-body potential that generates a good trial wave function for projecting
a large scale original Hamiltonian to a local subsystem with a small number of basis.
The resultant ground state of the projected Hamiltonian can locally approximate the true ground state.
However,
the lack of the variational principle makes it difficult to judge the quality of the choice of the potential.
Here we focus on the entanglement spectrum (ES) as a judging criterion;
accurate evaluation of the ES guarantees that
the corresponding reduced density matrix well reproduces all physical quantities on the local subsystem. 
We apply the DMET to the Hubbard model on the one-dimensional chain, zigzag chain, and triangular lattice and test several variants of potentials and cost functions. 
It turns out that ES serves as a more sensitive quantity than the energy and double occupancy 
to probe the quality of the DMET outcomes.
A symmetric potential reproduces the ES of the phase that continues from a noninteracting limit. 
The Mott transition as well as symmetry-breaking transitions can be detected by the singularities in the ES.
However, the details of the ES in the strongly interacting parameter
region depends much on these variants, meaning that 
the present DMET algorithm allowing for numerous variant is insufficient 
to fully characterize the particular phases that require characterization by the ES.
\end{abstract}

% title
\maketitle

%%%%%%%%%%%%%%%%%%%%%%%%%%%%%%%%%%%%%%%%%%%%
%%%%%%%%%%%%%%% introduction %%%%%%%%%%%%%%%
%%%%%%%%%%%%%%%%%%%%%%%%%%%%%%%%%%%%%%%%%%%%
\section{introduction}
\label{sec:introduction}
Quantum entanglement is recognized as one of the key modern concepts in quantum many-body physics.
In strongly correlated systems, it is an important theoretical tool to characterize several classes of phases of matter.
From the scaling behavior of entanglement entropy (EE),
one can extract the central charge associated with the underlying conformal field theory \cite{holzhey1994npb,vidal2003prl,calabrese2004jsm}.
The residual constant term in the EE is used to detect the existence of
a so-called topological order characterized by the long-range nature of the entanglement \cite{hamma2005pra,kitaev2006prl,levin2006prl}.
The entanglement spectrum (ES) is useful to identify a more subtle type of topological properties like
symmetry-protected topological phases \cite{gu2009prb,pollmann2010prb,fidkowski2011prb,turner2011prb,chen2011prb},
and its low-lying levels reflect the edge modes of the system \cite{li2008prl,fidkowski2010prl,turner2010prb,sterdyniak2012prb}.
While the entanglement-related properties turned out to serve as a probe for
such unusual phases beyond the Landau's symmetry-breaking paradigm,
they are also very useful to detect conventional quantum phase transitions
as it contains information on how the quantum many-body wave function is constructed from the local basis
\cite{deng2006prb,legeza2006prl,song2006prb,lou2006prb,ren2009pra,frerot2016prl}.
\par
To compute the ES or EE, however, 
one needs to evaluate many-body wave functions with enough accuracy for a large enough system size. 
This is generally difficult for strongly correlated two-dimensional (2D) many-body systems 
which are the platform of the above-mentioned exotic phenomena.
So far, numbers of numerical solvers have been applied or newly developed 
such as the quantum and variational Monte Carlo simulation \cite{suzuki1976ptp,foulkes2001rmp,gros1988prb,giamarchi1991prb,dev1992prb},
dynamical mean-field theory (DMFT) \cite{metzner1989prl,georges1992prb,georges1996rmp},
density matrix renormalization group (DMRG) \cite{white1992prl,white1993prb,schollwock2011ap},
and tensor network method \cite{nishio2004arxiv,jordan2008prl,orus2014ap}. 
They, however, suffer from the negative sign problems, lack of long-range quantum fluctuation, or
the large numerical cost because of the area law of the EE.
The further improvements of existing approaches as well as developments of new methods are desired.
\par
In the present paper, we focus on the density matrix embedding theory (DMET) introduced quite recently in Ref. [\onlinecite{knizia2012prl}].
The method tries to find the ground-state properties of fermionic lattice models
with low computational cost, and has been applied to various problems
\cite{knizia2012prl,chen2014prb,bulik2014prb,fan2015prb,booth2015prb,sandhoefer2016prb,zheng2016prb,wouters2017book,
zheng2017prb,mukherjee2017prb,gunst2017prb,ayral2017prb,fertitta2018prb,zheng2018thesis,lee2019prb,senjean2019prb,wu2020prb}
including quantum chemistry problems
\cite{knizia2013jctc,tsuchimochi2015jcp,wouters2016jctc,fulde2017jcp,kretchmer2018jcp,pham2018jctc,ye2018jcp,hermes2019jctc,reinhard2019jctc,cui2020jctc,wu2019jcp}.
It is also extended to the study of finite-temperature and dynamical properties \cite{booth2015prb,kretchmer2018jcp,sun2020prb}.
The key idea is to divide a large system into a small subsystem called an impurity fragment and a rest,
and to represent the original Hamiltonian by the small number of basis set
consisting of those inside the impurity and those selected from outside.
To properly select the basis set,
a noninteracting reference Hamiltonian with a one-body potential is prepared.
Among the constituent of its ground state wave function,
the local basis belonging to the impurity and those outside but entangled with the impurity basis
are chosen to represent the original Hamiltonian.
If this choice is correct,
the projected impurity Hamiltonian will yield the state
that reproduces well the true ground state of the original Hamiltonian in the impurity region \cite{knizia2012prl}.
\par
Although the original paper of the DMET by Knizia and Chan
refer to the DMFT as the similar cluster method \cite{knizia2012prl},
its construction is completely different;
the DMFT focuses not on the wave function but on the frequency-dependent Green's function \cite{metzner1989prl,georges1992prb,georges1996rmp},
and does not take account of the longer-range fluctuation or correlation effect.
The DMET is not a typical cluster method since the correlation functions inside the cluster does not depend much on the choices
of its shape \cite{xavier2020arxiv}, unlike the cluster-DMFT.
From that context, the DMET is rather built on the similar concept with
the first principles density functional theory (DFT),
%\textit{ab initio} first-principles method,
which assumes the existence of the one-body potential
that exactly reproduces the spatial distribution of the charge density of the true ground state
based on the Kohn-Sham theory,
and obtains the corresponding wave function based on the variational principle \cite{hohenberg1964pr,kohn1965pr}.
The DMET also prepares the one-body potential that reproduces \textit{the local one-body density matrix},
which includes both the information on the charge density of the impurity sites as well as the other off-diagonal properties.
\par
There are several advantages of the DMET over cluster-DMFT and variational cluster methods \cite{potthoff2012book};
the results do not depend much on the shape of the impurity cluster \cite{xavier2020arxiv},
and show significant convergence with the increase of the size of the impurity \cite{zheng2017prb}.
In fact,
the superconducting and stripe phases of the two-dimensional Hubbard model are well described by the DMET \cite{zheng2016prb},
which was not easy or almost impossible for previous solvers such as Gutzwiller variational methods, DMFT, or even QMC.
\par
However, the obtained ground-state energy is \textit{not variational} \cite{knizia2012prl}, 
and resultantly, the optimization procedures are not straightforward. 
Although previous benchmark studies show highly-accurate ground-state energy and double occupancy
for the Hubbard model on the one-dimensional (1D) chain and square lattice \cite{knizia2012prl,bulik2014prb,senjean2019prb,wu2020prb},
this does not necessarily guarantee the high performance of the DMET for obtaining other physical quantities. 
Different types of one-body potentials generally provide different physical quantities,
but one cannot determine which type is the best.
\par
The purpose of this paper is to give a comparative study on several variants of the DMET,
where, on top of several potential advantages of the method presented by the previous studies,
we clarify the underlying difficulty of reliably disclosing the deep quantum structure of the wave function when the system becomes strongly correlated.
To judge the quality of the potential,
we focus on the ES,
which has a one-to-one correspondence with eigenvalues of the reduced density matrix of the subsystem.
If the DMET properly reproduces the exact ES,
the reduced density matrix gives all the local physical quantities of the subsystem \cite{footnote1}.
Previously in an interacting spinless fermionic model, the density embedding theory (DET), one of the simplest variants of
the DMET, is shown to reproduce well the exact ES and to detect the phase transition \cite{xavier2020arxiv}.
Here we further examine the ES systematically by the DET and DMET with various types of the one-body potential
for the Hubbard model on the 1D chain, zigzag chain, and triangular lattice.
By comparing the DMET and DMRG results of the 1D chain and zigzag chain,
we find that the symmetric one-body potential well reproduces the low-level ES in the phase that continues from the noninteracting limit,
and in other phases the DMET fails to describe the exact ES.
We also find that the proper symmetry-breaking one-body potential leads to the singularity of the ES,
which is a good indicator of phase transitions.
However, we need to test several variants,
since the transition points may or may not appear depending on the choice of the potentials.
Furthermore,
the better agreement of the ground-state energy does not necessarily match the accuracy of the ES.
To show the reliability of the solution for unexplored models or exotic phases of matter,
the examination of ES in addition to the energy is required.
The paper aims to disclose these aspects of the DMET through its application to the frustrated triangular lattice model 
which typically hosts such exotic phases.
\par
This paper is organized as follows.
In Section \ref{sec:dmet},
we first review the basic algorithm of the DMET and discuss several variants of the one-body potential.
In Section \ref{sec:results},
we apply the DMET to the Hubbard model
defined on the 1D chain, zigzag chain, and triangular lattice,
and show their ground-state energy, double occupancy, ES, and EE.
We then give a conclusion in Section \ref{sec:conclusion}.

%%%%%%%%%%%%%%%%%%%%%%%%%%%%%%%%%%%%%%%%%%%%%%%%
%%%%%%%%%%%%%%% model and method %%%%%%%%%%%%%%%
%%%%%%%%%%%%%%%%%%%%%%%%%%%%%%%%%%%%%%%%%%%%%%%%
\section{density matrix embedding theory}
\label{sec:dmet}

So far,
the details of the method are fragmentally modified and tested from the original paper \cite{knizia2012prl,knizia2013jctc,bulik2014prb,wouters2017book},
e.g. the choice of the impurity basis representation,
the order of optimization,
whether to include the interaction of a bath exactly or in a mean-field form.
After examining them,
we selected the optimal algorithm which we explain in this section.
Then we introduce several types of the one-body potential we adopt shortly.

%%%%%%%%%% formulation %%%%%%%%%%
\subsection{Formulation}
\label{subsec:formulation}

We deal with the half-filled Hubbard model consisting of $N$ sites,
%%%%%%%%%%%%%%%
\begin{equation}
\hat{\mathcal{H}}
=
\sum_{i,j=1}^{N}
\sum_{\sigma}
t_{i,j}
\hat{c}_{i,\sigma}^{\dagger}
\hat{c}_{j,\sigma}
+
U
\sum_{i=1}^{N}
\hat{n}_{i,\uparrow}
\hat{n}_{i,\downarrow}
,
\label{eq:H}
\end{equation}
%%%%%%%%%%%%%%%
where
$t_{i,j}=t_{j,i}\in\mathbb{R}$ is the hopping amplitude between $i$ and $j$ sites,
$U$ is the on-site interaction,
$\hat{c}_{i,\sigma}^{\dagger}$ ($\hat{c}_{i,\sigma}$) denotes the creation (annihilation) of a fermion at site $i$ with spin $\sigma=\uparrow,\downarrow$,
and $\hat{n}_{i,\sigma}=\hat{c}_{i,\sigma}^{\dagger}\hat{c}_{i,\sigma}$.
We assume the hopping amplitude as $t_{i,j}=-t$ for nearest-neighboring $i$ and $j$ sites,
where the sign of $t$ may change at the boundary,
and $t_{i,j}=0$ for others.
We adopt an antiperiodic boundary condition which lifts the degeneracy of the one-body 
energy of the trial states for a given potential. 
\par
The core process of the DMET is to construct an impurity model using the Schmidt decomposition of a trial state $\ket{\Psi}$ \cite{knizia2012prl}.
We divide the entire system into two subsystems A and B consisting of $N_{\mathrm{A}}$ and $N_{\mathrm{B}}=N-N_{\mathrm{A}}$ sites,
and suppose $N_{\mathrm{A}}\ll N_{\mathrm{B}}$.
Here the subsystem A corresponds to an impurity fragment. 
Then $\ket{\Psi}$ can be written as
%%%%%%%%%%%%%%%
\begin{equation}
\ket{\Psi}
=
\sum_{n=1}^{\chi}
\lambda_{n}(\Psi)
\ket{\Psi_{n}^{[\mathrm{A}]}}
\otimes
\ket{\Psi_{n}^{[\mathrm{B}]}}
,
\label{eq:schmidt}
\end{equation}
%%%%%%%%%%%%%%%
where $\chi=4^{N_{\mathrm{A}}}$,
$\lambda_{n}(\Psi)\geq0$ is the Schmidt coefficient,
and $\{\ket{\Psi_{n}^{[X]}}\}_{n=1}^{\chi}$ forms an orthonormal basis in the subsystem $X$=A, B.
Here, the DMET takes advantage of the fact that only a small number of basis 
from among those belonging to the large subsystem B is required to describe the trial state: 
the set $\{\ket{\Psi_{n}^{[\mathrm{B}]}}\}_{n=1}^{\chi}$ is called bath state \cite{knizia2012prl}.
The process is completed by applying a projection operator 
%%%%%%%%%%%%%%%
\begin{equation}
\hat{\mathcal{P}}
=
\hat{1}^{[\mathrm{A}]}
\otimes
\left(
\sum_{n=1}^{\chi}
\ket{\Psi_{n}^{[\mathrm{B}]}}
\bra{\Psi_{n}^{[\mathrm{B}]}}
\right)
,
\label{eq:P}
\end{equation}
%%%%%%%%%%%%%%%
where $\hat{1}^{[\mathrm{A}]}$ is the identity operator defined on the subsystem A.
The entire system is then reduced to the impurity fragment (subsystem A) coupled to an external bath,
and the original Hamiltonian is transformed to an impurity Hamiltonian,
%%%%%%%%%%%%%%%
\begin{equation}
\hat{\mathcal{H}}_{\mathrm{imp}}
=
\hat{\mathcal{P}}\hat{\mathcal{H}}\hat{\mathcal{P}}
.
\label{eq:Himp}
\end{equation}
%%%%%%%%%%%%%%%
Note that the number of bath states is equal to $4^{N_{\mathrm{A}}}$, 
and accordingly the dimension of the Hilbert space to be searched for is 
significantly reduced from the original one.
One can thus obtain the ground state of the impurity Hamiltonian, $\ket{\Phi_{\mathrm{imp}}}$,
using a high-accuracy numerical method such as the exact diagonalization. 
%%%%%%%%%%%%%%%%%%%%%%%%%%%%%%%%%%%%%%%%%%%%%%%%%%%%%%%%%%%%%%%%%%%%%%%%%%%%%%%%%%%%%%%%%%%%%%%%%%%%%%%%%%%%%%%%%%%%%%%%
\begin{figure}[t]
\includegraphics[width=85mm]{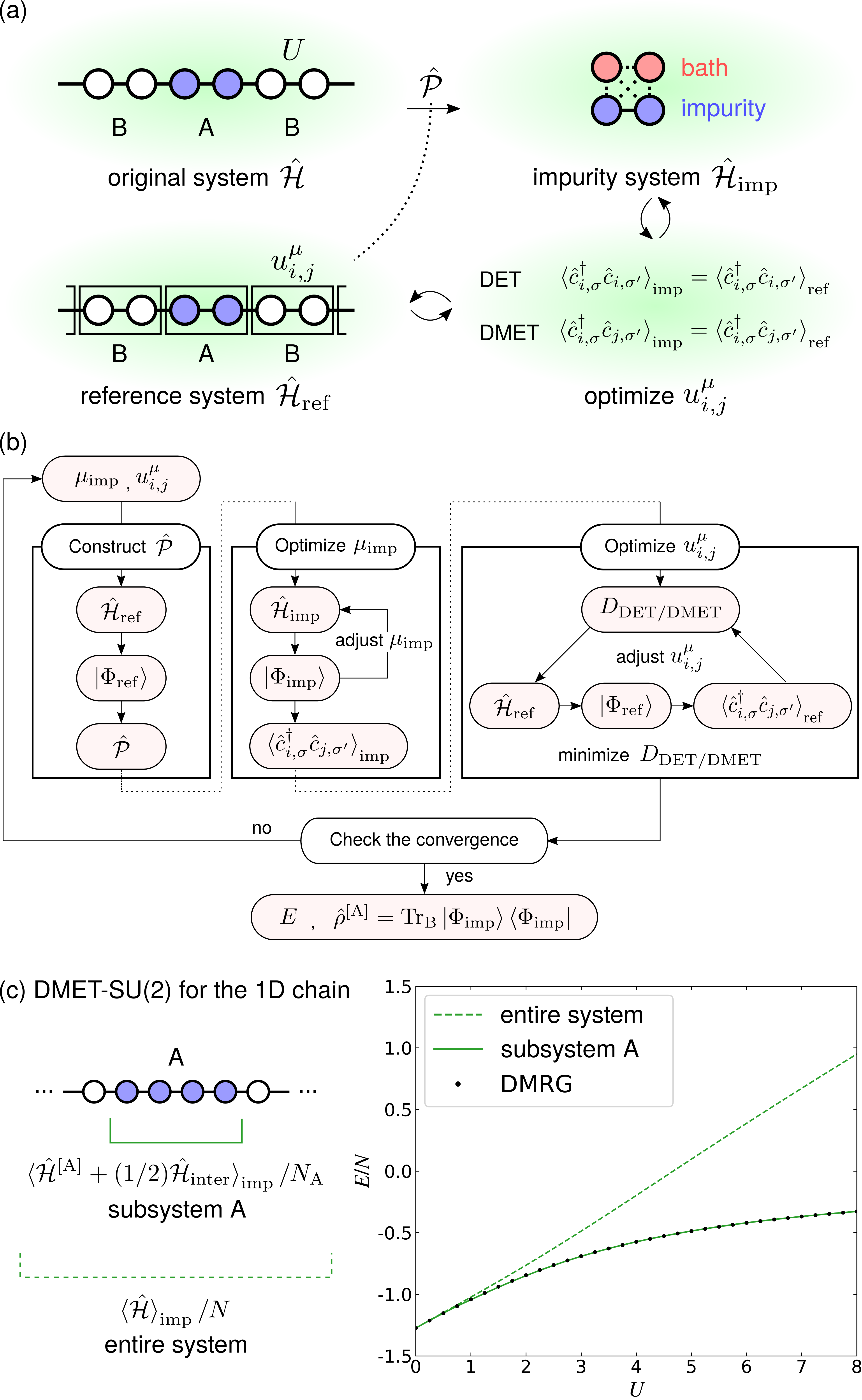}
\caption{(a) Schematic illustration and (b) flowchart of the DET and DMET scheme.
The original system is mapped to the impurity one by the projection operator,
which is constructed from the Schmidt decomposition of the ground state of the reference system.
The one-body potential in the reference system is optimized
such that the one-body density or density matrix of the reference system matches that of the impurity one.
(c) Energy per site as the function of $U$.
The energy density of the entire system deviates from the DMRG result as increasing $U$ while that of subsystem A does not.}
\label{fig1}
\end{figure}
%%%%%%%%%%%%%%%%%%%%%%%%%%%%%%%%%%%%%%%%%%%%%%%%%%%%%%%%%%%%%%%%%%%%%%%%%%%%%%%%%%%%%%%%%%%%%%%%%%%%%%%%%%%%%%%%%%%%%%%%
\par
The ground state $\ket{\Phi_{\mathrm{imp}}}$ of Eq. (\ref{eq:Himp}) depends much on the choice of the trial state. 
The best choice of the trial state is the exact ground state, in which case 
$\ket{\Phi_{\mathrm{imp}}}$ matches the exact ground state by definition. 
However, the exact solution is \textit{a priori} unknown. 
The approximate trial state shall better reproduce the entanglement between the impurity and the bath of the exact ground state, 
and is prepared as the ground state of the following reference Hamiltonian \cite{knizia2012prl},
%%%%%%%%%%%%%%%
\begin{equation}
\hat{\mathcal{H}}_{\mathrm{ref}}
=
\sum_{i,j=1}^{N}
\sum_{\sigma}
t_{i,j}
\hat{c}_{i,\sigma}^{\dagger}
\hat{c}_{j,\sigma}
+
\sum_{i,j=1}^{N}
\hat{\bm{c}}_{i}^{\dagger}
(
u_{i,j}^{0}\sigma^{0}
+
\bm{u}_{i,j}\cdot\bm{\sigma}
)
\hat{\bm{c}}_{j}
,
\label{eq:Href}
\end{equation}
%%%%%%%%%%%%%%%%%%%%%%%%%%%%%%%%%%%%%%%%%%%%%
where $u_{i,j}^{0}$ and $\bm{u}_{i,j}=(u_{i,j}^{x},u_{i,j}^{y},u_{i,j}^{z})$ with $u_{i,j}^{\mu}=(u_{j,i}^{\mu})^{*}$ ($\mu=0,x,y,z$)
are the one-body potential puts by hand, 
$\sigma^{0}$ and $\bm{\sigma}=(\sigma^{x},\sigma^{y},\sigma^{z})$ is the unit and Pauli matrices, 
and $\hat{\bm{c}}_{i}=(\hat{c}_{i,\uparrow},\hat{c}_{i,\downarrow})^{T}$.
We will discuss several types of one-body potentials in Subsection \ref{subsec:symmetry}.
The ground state of the reference Hamiltonian, $\ket{\Psi}=\ket{\Phi_{\mathrm{ref}}}$, 
is written as a Slater determinant, 
which is analytically decomposed \cite{israel2006jpa,peschel2012bjp} into the form as Eq. (\ref{eq:schmidt}).
The basis in the subsystem A, $\ket{\Psi_{\bm{n}}^{[\mathrm{A}]}}$, 
is not included in the projection operator (\ref{eq:P}) 
since the original fermionic operator $\hat{c}_{i,\sigma}$ ($i=1,2,\cdots,N_{\mathrm{A}}$) is fully preserved 
and the impurity Hamiltonian in the subsystem A is not modified.
On the other hand,
the projection operator (\ref{eq:P}) depends on the bath state $\ket{\Psi_{\bm{n}}^{[\mathrm{B}]}}$,
which is given by \cite{israel2006jpa,peschel2012bjp}
%%%%%%%%%%%%%%%
\begin{equation}
\ket{\Psi_{\bm{n}}^{[\mathrm{B}]}}
=
\prod_{\alpha=1}^{2N_{\mathrm{A}}}
(\hat{b}_{\alpha}^{\dagger})^{n_{\alpha}}
\left(
\prod_{\beta=1}^{N-2N_{\mathrm{A}}}
\hat{e}_{\beta}^{\dagger}
\right)
\ket{0^{[\mathrm{B}]}}
,
\label{eq:Psi_B}
\end{equation}
%%%%%%%%%%%%%%%
where $\bm{n}=(n_{1},\cdots,n_{2N_{\mathrm{A}}})$ with $n_{\alpha}=0,1$,
$\ket{0^{[\mathrm{B}]}}$ is the vacuum on the subsystem B,
and the operators $\hat{b}_{\alpha}$ and $\hat{e}_{\beta}$ are called bath and core orbitals,
which live on the subsystem B.
These orbitals are obtained by the unitary transformation of the site-based one $\hat{c}_{i,\sigma}$ as \cite{wouters2017book}
%%%%%%%%%%%%%%%
\begin{equation}
\hat{\bm{c}}_{\mathrm{B}}
=
L_{\mathrm{B}}
\begin{pmatrix}
\hat{\bm{b}}\\
\hat{\bm{e}}\\
\hat{\bm{f}}
\end{pmatrix}
,
\end{equation}
%%%%%%%%%%%%%%%
where
$\hat{\bm{c}}_{\mathrm{B}}=(\hat{c}_{N_{\mathrm{A}}+1,\uparrow},\hat{c}_{N_{\mathrm{A}}+1,\downarrow},\cdots,\hat{c}_{N,\uparrow},\hat{c}_{N,\downarrow})^{T}$,
is the set of fermions on the subsystem B,
$\hat{\bm{b}}=(\hat{b}_{1},\cdots,\hat{b}_{2N_{\mathrm{A}}})^{T}$,
$\hat{\bm{e}}=(\hat{e}_{1},\cdots,\hat{e}_{N-2N_{\mathrm{A}}})^{T}$,
and
$\hat{\bm{f}}=(\hat{f}_{1},\cdots,\hat{f}_{N-2N_{\mathrm{A}}})^{T}$
is the remaining orbitals which do not appear in the bath state (\ref{eq:Psi_B}).
The $2N_{\mathrm{B}}\times2N_{\mathrm{B}}$ unitary matrix $L_{\mathrm{B}}$ is obtained by the singular value decomposition of the eigen wave function of Eq. (\ref{eq:Href})
and is improved by optimizing the one-body potential \cite{wouters2017book}.
The operator $\hat{c}_{i,\sigma}$ on the subsystem B is then transformed as
the linear combination of the bath, core, and remaining orbitals,
and after the projection by using the bath state (\ref{eq:Psi_B}),
the kinetic and interaction terms of the bath orbitals $\hat{\bm{b}}$ are reflected exactly in the impurity Hamiltonian,
while the core orbitals $\hat{\bm{e}}$ are treated as mean-field potentials and the remaining orbitals $\hat{\bm{f}}$ are discarded.
\par
The impurity Hamiltonian (\ref{eq:Himp}) is explicitly given as
%%%%%%%%%%%%%%%%%%%%%%%%%%%%%%%%%%%%%%%%%%%%%
\begin{equation}
\hat{\mathcal{H}}_{\mathrm{imp}}
=
\hat{\mathcal{H}}^{[\mathrm{A}]}
+
\hat{\mathcal{H}}_{\mathrm{bath}}
+
\hat{\mathcal{H}}_{\mathrm{inter}}
,
\end{equation}
%%%%%%%%%%%%%%%
%%%%%%%%%%%%%%%
\begin{equation}
\hat{\mathcal{H}}^{[\mathrm{A}]}
=
\sum_{i,j=1}^{N_{\mathrm{A}}}
\sum_{\sigma}
t_{i,j}
\hat{c}_{i,\sigma}^{\dagger}
\hat{c}_{j,\sigma}
+
U
\sum_{i=1}^{N_{\mathrm{A}}}
\hat{n}_{i,\uparrow}
\hat{n}_{i,\downarrow}
,
\end{equation}
%%%%%%%%%%%%%%%
%%%%%%%%%%%%%%%
\begin{equation}
\hat{\mathcal{H}}_{\mathrm{bath}}
=
\sum_{\alpha,\beta=1}^{2N_{\mathrm{A}}}
t_{\alpha,\beta}^{(\mathrm{bath})}
\hat{b}_{\alpha}^{\dagger}
\hat{b}_{\beta}
+
\sum_{\alpha,\beta,\gamma,\delta=1}^{2N_{\mathrm{A}}}
U_{\alpha,\beta,\gamma,\delta}^{(\mathrm{bath})}
\hat{b}_{\alpha}^{\dagger}
\hat{b}_{\beta}^{\dagger}
\hat{b}_{\gamma}
\hat{b}_{\delta}
,
\label{eq:Hbath}
\end{equation}
%%%%%%%%%%%%%%%
%%%%%%%%%%%%%%%
\begin{equation}
\hat{\mathcal{H}}_{\mathrm{inter}}
=
\sum_{i=1}^{N_{\mathrm{A}}}
\sum_{\sigma}
\sum_{\alpha=1}^{2N_{\mathrm{A}}}
\left(
t_{(i,\sigma),\alpha}^{(\mathrm{inter})}
\hat{c}_{i,\sigma}^{\dagger}
\hat{b}_{\alpha}
+
\mathrm{h.c.}
\right)
,
\label{eq:Hinter}
\end{equation}
%%%%%%%%%%%%%%%
where $\hat{\mathcal{H}}^{[\mathrm{A}]}$ is the original Hamiltonian on the subsystem A,
$\hat{\mathcal{H}}_{\mathrm{bath}}$ is the Hamiltonian on the bath,
and $\hat{\mathcal{H}}_{\mathrm{inter}}$ is the hopping terms between the impurity and bath sites.
Here we omit the constant term in $\hat{\mathcal{H}}_{\mathrm{bath}}$.
%The operator $\hat{b}_{\alpha}$ ($\alpha=1,2,\cdots,2N_{\mathrm{A}}$) belongs to the bath orbital, 
%which lives on the subsystem B
%and is obtained by the unitary transformation of the site-based one $\hat{c}_{i,\sigma}$ \cite{wouters2017book}.
%The redundancy of the number of orbitals on the subsystem B is treated as such that 
%the kinetic and interaction terms of the bath orbitals are reflected exactly in the impurity Hamiltonian, 
%and the remaining $(2N_{\mathrm{B}}-2N_{\mathrm{A}})$-orbitals are either discarded 
%or treated as mean-field potentials included in $t_{\alpha,\beta}^{(\mathrm{bath})}$.
%The hopping amplitudes and two-body interaction in Eqs. (\ref{eq:Hbath}) and (\ref{eq:Hinter}) are determined
%from the matrix elements $\braket{\Psi_{n}^{[\mathrm{B}]}|\hat{\mathcal{H}}|\Psi_{m}^{[\mathrm{B}]}}$ \cite{knizia2012prl}.
The hopping amplitudes and two-body interaction in Eqs. (\ref{eq:Hbath}) and (\ref{eq:Hinter}) are determined
from the unitary matrix $L_{\mathrm{B}}$ \cite{wouters2017book}.
Here,
$\hat{\mathcal{H}}_{\mathrm{bath}}$ includes the four-fermionic interaction term because we assume the original Hamiltonian to be two-body.
In general,
the $n$-body term of the original Hamiltonian is transformed to the $2n$-fermionic interaction terms consisting of several different indices of $\hat{b}$'s.
\par
The remaining issue is an optimization of the one-body potential.
In the DMET, 
the one-body potential is chosen as such that the one-body density matrix obtained by the ground state of the reference Hamiltonian  
matches the one-body density matrix of the ground state of the impurity Hamiltonian \cite{knizia2012prl},
%%%%%%%%%%%%%%%
\begin{equation}
\braket{\hat{c}_{i,\sigma}^{\dagger}\hat{c}_{j,\sigma'}}_{\mathrm{imp}}
=
\braket{\hat{c}_{i,\sigma}^{\dagger}\hat{c}_{j,\sigma'}}_{\mathrm{ref}}
\end{equation}
%%%%%%%%%%%%%%%
for $i,j=1,2,\cdots,N_{\mathrm{A}}$ and $\sigma,\sigma'=\uparrow,\downarrow$,
where
$\braket{\hat{\mathcal{O}}}_{\mathrm{imp}}=\braket{\Phi_{\mathrm{imp}}|\hat{\mathcal{O}}|\Phi_{\mathrm{imp}}}$
and
$\braket{\hat{\mathcal{O}}}_{\mathrm{ref}}=\braket{\Phi_{\mathrm{ref}}|\hat{\mathcal{O}}|\Phi_{\mathrm{ref}}}$
for an operator $\hat{\mathcal{O}}$. 
%The \textit{ab initio} first-principles method,
The first principles DFT,
whose guiding principle shall be compared to the DMET,
assumes the existence of a one-body potential
that reproduces the same charge distribution of the correlated many-body wave function by the noninteracting one.
Similarly, the present one-body potential is expected to have the one-body density matrix that mimics the correlated ones. 
The measure of optimization is the cost function given as 
%%%%%%%%%%%%%%%
\begin{equation}
D_{\mathrm{DMET}}
=
\sum_{i,j=1}^{N_{\mathrm{A}}}
\sum_{\sigma,\sigma'}
\left|
\braket{\hat{c}_{i,\sigma}^{\dagger}\hat{c}_{j,\sigma'}}_{\mathrm{imp}}
-
\braket{\hat{c}_{i,\sigma}^{\dagger}\hat{c}_{j,\sigma'}}_{\mathrm{ref}}
\right|^{2}
.
\label{eq:D_DMET}
\end{equation}
Minimizing $D_{\mathrm{DMET}}$ approximately maximizes the overlap between
the one-body density matrix of the reference system and the impurity one.
\par
The simpler variant of the DMET is also proposed in Ref. [\onlinecite{bulik2014prb}], called DET,
in which the one-body potential is optimized in order to satisfy
%%%%%%%%%%%%%%%
\begin{equation}
\braket{\hat{c}_{i,\sigma}^{\dagger}\hat{c}_{i,\sigma'}}_{\mathrm{imp}}
=
\braket{\hat{c}_{i,\sigma}^{\dagger}\hat{c}_{i,\sigma'}}_{\mathrm{ref}}
,
\label{eq:det}
\end{equation}
%%%%%%%%%%%%%%%
for $i=1,2,\cdots,N_{\mathrm{A}}$ and $\sigma,\sigma'=\uparrow,\downarrow$.
Equation (\ref{eq:det}) guarantees an exact fitting of a particle and spin density between the impurity system and the reference system. 
The corresponding cost function is
%%%%%%%%%%%%%%%
\begin{equation}
D_{\mathrm{DET}}
=
\sum_{i=1}^{N_{\mathrm{A}}}
\sum_{\sigma,\sigma'}
\left|
\braket{\hat{c}_{i,\sigma}^{\dagger}\hat{c}_{i,\sigma'}}_{\mathrm{imp}}
-
\braket{\hat{c}_{i,\sigma}^{\dagger}\hat{c}_{i,\sigma'}}_{\mathrm{ref}}
\right|^{2}
,
\label{eq:D_DET}
\end{equation}
%%%%%%%%%%%%%%%
and $D_{\mathrm{DET}}=0$ leads to Eq. (\ref{eq:det}).
Schematic illustration of the DET and DMET scheme is shown in Fig. \ref{fig1}(a).
\par
Let us add some remarks on further details of the calculation. 
In the DMET, the particle density in the impurity region, 
$\sum_{i=1}^{N_{\mathrm{A}}}\sum_{\sigma}\braket{\hat{n}_{i,\sigma}}_{\mathrm{imp}}/N_{\mathrm{A}}$,
does not necessarily equal the correct particle density $1.0$.
To solve this issue, a fictitious chemical potential $\mu_{\mathrm{imp}}$ is introduced in the impurity region as \cite{wouters2016jctc}
%%%%%%%%%%%%%%%
\begin{equation}
\hat{\mathcal{H}}_{\mathrm{imp}}
\to
\hat{\mathcal{H}}_{\mathrm{imp}}
-
\mu_{\mathrm{imp}}
\sum_{i=1}^{N_{\mathrm{A}}}
\hat{n}_{i}
,
\end{equation}
%%%%%%%%%%%%%%%
where $\mu_{\mathrm{imp}}$ is determined by 
adjusting the particle density in the impurity region to $1.0$.
Another detail is how to minimize the cost function. 
Calculating its derivative with respect to the one-body potential 
requires large computational cost since one needs to solve the impurity model many times in 
evaluating the derivative of $\braket{\hat{c}_{i,\sigma}^{\dagger}\hat{c}_{j,\sigma'}}_{\mathrm{imp}}$. 
This process is avoided by performing the minimization by a self-consistent procedure,
whose detail is explained in Ref. [\onlinecite{wouters2017book}].
We use either the Broyden-Fietcher-Goldfarb-Shanno (BFGS) algorithm or the Powell algorithm instead \cite{fletcher1987},
depending on the band structure of the reference system.
The self-consistent loop is iterated until the change of $\mu_{\mathrm{imp}}$ and the maximum value of $u_{i,j}^{\mu}$ is below $5\times10^{-4}$.
We show the flowchart of the DMET (DET) calculation in Fig. \ref{fig1}(b).

%%%%%%%%%% formulation %%%%%%%%%%
\subsection{Reduced density matrix}
\label{subsec:rdm}
Once the values of the chemical potential and one-body potential are converged,
the ground state of the impurity Hamiltonian $\ket{\Phi_{\mathrm{imp}}}$ can be obtained.
Here, we point out that $\ket{\Phi_{\mathrm{imp}}}$ itself is \textit{totally different} from the true ground state.
We show in Fig. \ref{fig1}(c) the energy densities of the Hubbard model on the 1D chain obtained
by using the entire wave function and by using only the local subsystem A, and compare them with the DMRG result.
Here the energy of the entire wave function is given by $\braket{\hat{\mathcal{H}}}_{\mathrm{imp}}$,
and that of the local subsystem A can be evaluated as
$\braket{\hat{\mathcal{H}}^{[\mathrm{A}]}+(1/2)\hat{\mathcal{H}}_{\mathrm{inter}}}_{\mathrm{imp}}$,
where the factor $1/2$ is introduced to avoid double counting.
We see that the energy density of the entire wave function significantly deviates from the DMRG result,
while that of the local subsystem A is in good agreement with it. 
The DMET optimizes the bath and the core states that represents the correlation between the impurity
and the rest of the system,
and make the impurity part of $\ket{\Phi_{\mathrm{imp}}}$ equivalent to that of the true ground state \cite{knizia2013jctc}.
\par
The accuracy of $\ket{\Phi_{\mathrm{imp}}}$ is physically in one-to-one correspondence with
the accuracy of the reduced density matrix 
%%%%%%%%%%%%%%%
\begin{equation}
\hat{\rho}^{[\mathrm{A}]}
=
\mathrm{Tr}_{\mathrm{B}}
\ket{\Phi_{\mathrm{imp}}}
\bra{\Phi_{\mathrm{imp}}}
, 
\end{equation}
%%%%%%%%%%%%%%%
since all the local physical quantities on the subsystem A 
represented by the operators $\hat{\mathcal{O}}^{[\mathrm{A}]}$ can be evaluated as
%%%%%%%%%%%%%%%
\begin{equation}
\braket{\hat{\mathcal{O}}^{[\mathrm{A}]}}_{\mathrm{imp}}
=
\mathrm{Tr}_{\mathrm{A}}
\left[
\hat{\rho}^{[\mathrm{A}]}
\hat{\mathcal{O}}^{[\mathrm{A}]}
\right]
.
\label{eq:Oimp}
\end{equation}
%%%%%%%%%%%%%%%
The measures to judge the quality of the reduced density matrix are the EE and ES; 
the EE $S_{\mathrm{A}}$ between the two subsystems is given by, 
%%%%%%%%%%%%%%%
\begin{equation}
S_{\mathrm{A}}
=
-
\mathrm{Tr}_{\mathrm{A}}
\left[
\hat{\rho}^{[\mathrm{A}]}
\ln\hat{\rho}^{[\mathrm{A}]}
\right]
,
\end{equation} 
%%%%%%%%%%%%%%%
and the ES, $\zeta_{n}$ ($n=1,2,\cdots,\chi$), is the set of eigenvalues of the entanglement Hamiltonian 
$-\ln\hat{\rho}^{[\mathrm{A}]}$ \cite{li2008prl}. 
\par
The ground state energy $E$ is an extensive quantity, 
and is evaluated by assuming that 
the entire system is tiled with the same impurity fragment \cite{wouters2017book} as
%%%%%%%%%%%%%%%
\begin{equation}
E
=
\left(
\frac{N}{N_{\mathrm{A}}}
\right)
\times
\Braket{\hat{\mathcal{H}}^{[\mathrm{A}]}+\frac{1}{2}\hat{\mathcal{H}}_{\mathrm{inter}}}_{\mathrm{imp}}
,
\end{equation}
%%%%%%%%%%%%%%%
where
$\braket{\hat{\mathcal{H}}^{[\mathrm{A}]}+(1/2)\hat{\mathcal{H}}_{\mathrm{inter}}}_{\mathrm{imp}}$ is
the energy of each impurity fragment as explained above.
Therefore, it would happen that $E$ including the term outside the impurity fragment
may become less accurate than the other quantities inside the impurity fragment
which are solely determined by $\hat{\rho}^{[\mathrm{A}]}$.
In this context, the ground state energy shall not be the good measure for the accuracy of the DMET.
%%%%%%%%%%%%%%%%%%%%%%%%%%%%%%%%%%%%%%%%%%%%%%%%%%%%%%%%%%%%%%%%%%%%%%%%%%%%%%%%%%%%%%%%%%%%%%%%%%%%%%%%%%%%%%%%%%%%%%%%
\begin{table}[t]
\begin{tabular}{|l|l|l|l|} \hline
type & potential & cost & symmetry \cite{footnote2}\\ \hline \hline % for revtex4-2
DET-SU(2) & $u_{i}^{0}\sigma^{0}$ & \multirow{2}{*}{$D_{\mathrm{DET}}$} & SU(2)\\ \cline{1-2} \cline{4-4}
DET-Full & $u_{i}^{0}\sigma^{0}+\bm{u}_{i}\cdot\bm{\sigma}$ & & break SU(2)\\ \hline
DMET-SU(2) & $u_{i,j}^{0}\sigma^{0}$ & \multirow{2}{*}{$D_{\mathrm{DMET}}$} & break TS\\ \cline{1-2} \cline{4-4}
DMET-Full & $u_{i,j}^{0}\sigma^{0}+\bm{u}_{i}\cdot\bm{\sigma}$ & & break TS and SU(2)\\ \hline
\end{tabular}
\caption{Four types of the one-body potential.
These potentials are defined in the impurity region and periodically repeated over the entire reference system.
TS is an acronym for translational symmetry.}
\label{table1}
\end{table}
%%%%%%%%%%%%%%%%%%%%%%%%%%%%%%%%%%%%%%%%%%%%%%%%%%%%%%%%%%%%%%%%%%%%%%%%%%%%%%%%%%%%%%%%%%%%%%%%%%%%%%%%%%%%%%%%%%%%%%%%

%%%%%%%%%% formulation %%%%%%%%%%
\subsection{Symmetry of the one-body potential}
\label{subsec:symmetry}
We examine several types of one-body potentials in the following. 
Let us first remark that the one-body potential generally breaks the translational symmetry
of the reference system by construction,
since the potential of the impurity region is repeated over the entire system,
i.e. $u_{i+N_{\mathrm{A}},j+N_{\mathrm{A}}}^{\mu}=u_{i,j}^{\mu}$ ($\mu=0,x,y,z$) \cite{wouters2017book}. 
The SU(2) spin-rotational symmetry is also not preserved for the general form of 
$\bm{u}_{i,j}$ in Eq. (\ref{eq:Href}).
With this in mind we deal with four types of one-body potentials,
and separately denote the DMET (DET) algorithm with these potentials
as DET-SU(2), DET-Full, DMET-SU(2) and DMET-Full, whose details are summarized in Table \ref{table1}.
In the DET-SU(2), 
the one-body potential has no off-diagonal terms and preserve the SU(2) spin-rotational symmetry,
i.e. $u_{i,j}^{0}=u_{i}^{0}\delta_{i,j}$ and $\bm{u}_{i,j}=\bm{0}$.
In this case,
the one-body potential is just a site-dependent chemical potential.
We see shortly that the optimized potential for 
the DET-SU(2) has a vanishingly small value of $u_{i}^{0}$ and does not break any symmetry of the reference system.
In the DET-Full, the one-body potential has a site-dependent ``magnetic field'',
$\bm{u}_{i,j}=\delta_{i,j}\bm{u}_{i}$,
which breaks the SU(2) spin-rotational symmetry.
For both the DET-SU(2) and DET-Full, we adopt $D_{\mathrm{DET}}$ as a cost function.
In the DMET-SU(2) and DMET-Full, 
the one-body potential has off-diagonal terms $u_{i,j}^{0}$ ($i\neq j$)
and is optimized by minimizing $D_{\mathrm{DMET}}$.
These off-diagonal elements can modify the kinetic term in the reference system.
Practically,
we allow only the intra-cluster hopping terms $u_{i,j}^{\mu}$ ($i,j=1,2,\cdots,N_{\mathrm{A}}$) which does not modify the results. 
This is reasonable because we need to keep the inter-cluster hopping finite in order to preserve the entanglement between the impurity and the bath,
and while at the same time, the unit of the energy can be chosen arbitrarily so that one of the parameters can be set to unity.
The degree of inter-cluster entanglement is relatively controlled by modifying $u_{i,j}^{\mu}$ inside the cluster.
The SU(2) spin-rotational symmetry is preserved in the DMET-SU(2) while not in the DMET-Full.
For practical reasons, 
we further assume
$\bm{u}_{i,j}=\bm{u}_{i}\delta_{i,j}$ and $\mathrm{Im}[u_{i,j}^{\mu}]=0$ ($\mu=0,x,y,z$) for all types,
which reduces the number of elements in $u_{i,j}^{\mu}$ and simplifies the optimization procedure. 
%%%%%%%%%%%%%%%%%%%%%%%%%%%%%%%%%%%%%%%%%%%%%%%%%%%%%%%%%%%%%%%%%%%%%%%%%%%%%%%%%%%%%%%%%%%%%%%%%%%%%%%%%%%%%%%%%%%%%%%%
\begin{figure}[t]
\includegraphics[width=85mm]{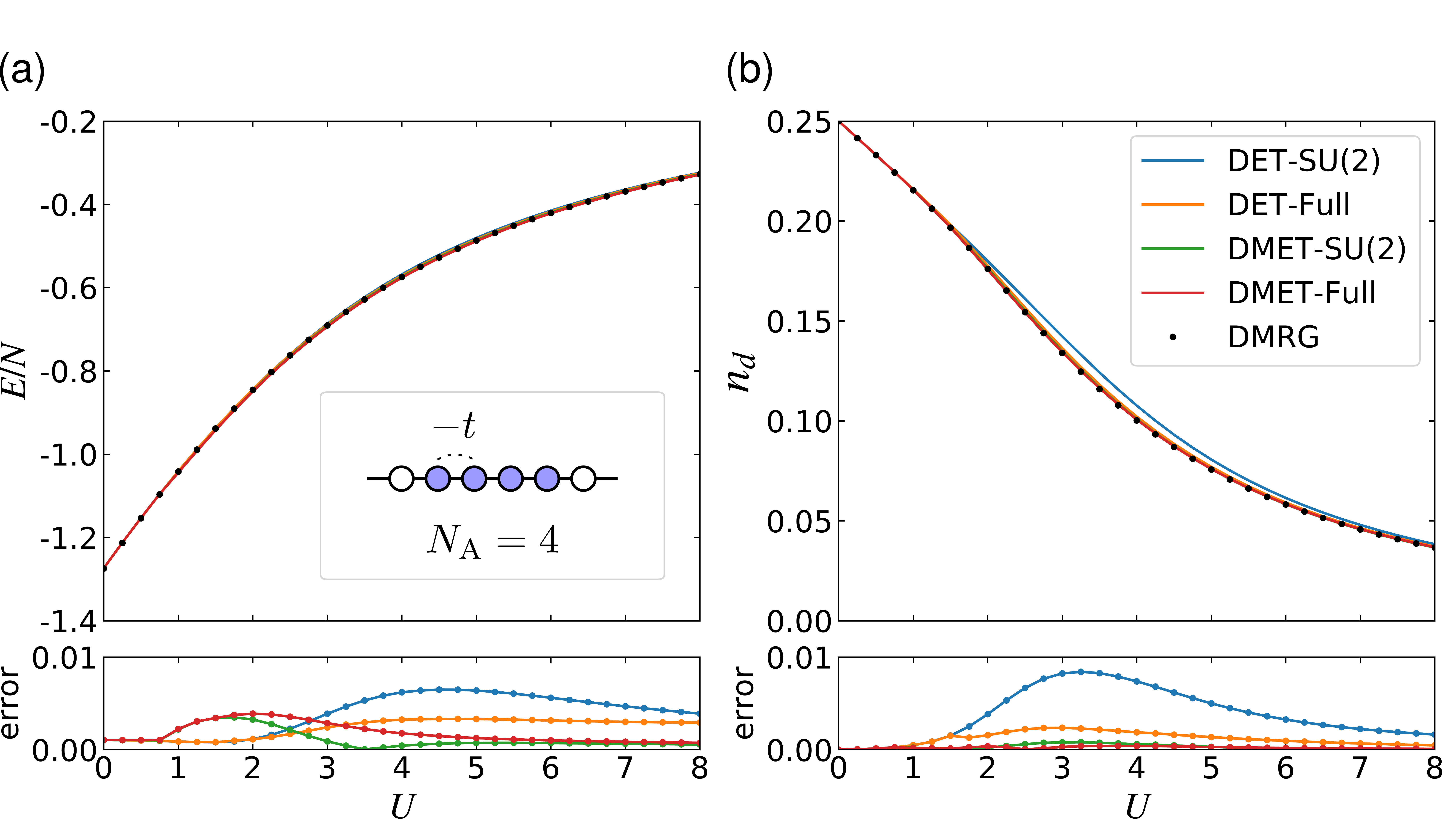}
\caption{(a) Ground-state energy per site and (b) double occupancy of the 1D Hubbard chain at half-filling
as the function of $U$ for $N=120$ and $N_{\mathrm{A}}=4$.
The bottom panel shows the error with respect to the DMRG results for $N=40$ and $m=800$ under the antiperiodic boundary condition.
All types reproduce the DMRG results with an insignificant error.}
\label{fig2}
\end{figure}
%%%%%%%%%%%%%%%%%%%%%%%%%%%%%%%%%%%%%%%%%%%%%%%%%%%%%%%%%%%%%%%%%%%%%%%%%%%%%%%%%%%%%%%%%%%%%%%%%%%%%%%%%%%%%%%%%%%%%%%%
%%%%%%%%%%%%%%%%%%%%%%%%%%%%%%%%%%%%%%%%%%%%%%%%%%%%%%%%%%%%%%%%%%%%%%%%%%%%%%%%%%%%%%%%%%%%%%%%%%%%%%%%%%%%%%%%%%%%%%%%
\begin{figure}[t]
\includegraphics[width=85mm]{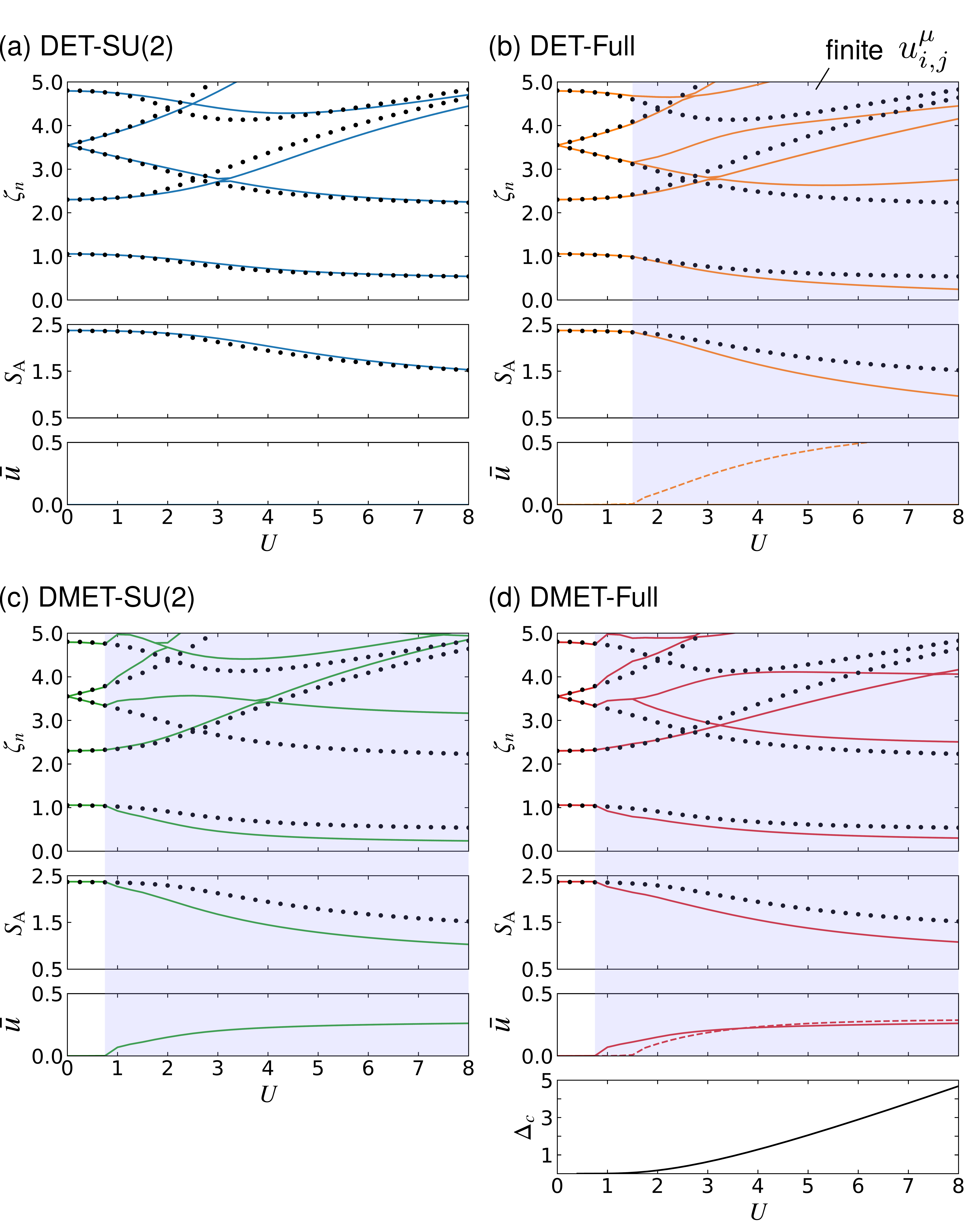}
\caption{ES, EE, and the average of the one-body potential of the 1D Hubbard chain at half-filling as the function of $U$
for (a) DET-SU(2), (b) DET-Full, (c) DMET-SU(2), and (d) DMET-Full.
Data points are the DMRG results.
The solid and dotted line in each bottom panel correspond to
the SU(2) symmetric ($\bar{u}^{0}$) and broken ($\sum_{\mu=x,y,z}\bar{u}^{\mu}$) part of the one-body potential, respectively.
The DET-SU(2),
in which the one-body potential is exactly zero,
well reproduces the overall structure of the exact low-energy ES and EE.
While in other types the ES and EE deviate from the DMRG results as the one-body potential develops,
which can be observed in the shaded region.
This deviation starts to develop when the charge gap $\Delta_{c}$ develops as well,
which is shown in the bottom panel of (d).
The charge gap $\Delta_{c}$ is obtaned by the Bethe ansatz.}
\label{fig3}
\end{figure}
%%%%%%%%%%%%%%%%%%%%%%%%%%%%%%%%%%%%%%%%%%%%%%%%%%%%%%%%%%%%%%%%%%%%%%%%%%%%%%%%%%%%%%%%%%%%%%%%%%%%%%%%%%%%%%%%%%%%%%%%

%%%%%%%%%%%%%%%%%%%%%%%%%%%%%%%%%%%%%%%
%%%%%%%%%%%%%%% results %%%%%%%%%%%%%%%
%%%%%%%%%%%%%%%%%%%%%%%%%%%%%%%%%%%%%%%
\section{results}
\label{sec:results}

In this section, we show the results obtained by the DET and DMET, 
where the exact diagonalization is used to find the ground state of the impurity model.
We set the nearest-neighbor hopping amplitude as unity, i.e. $t=1$.
Our results for the 1D chain and zigzag chain are compared with the DMRG data.

%%%%%%%%%%%%%%% Chain %%%%%%%%%%%%%%%
\subsection{1D chain}
\label{subsec:chain}

We first consider the Hubbard model on the 1D chain at half-filling.
This model is exactly solved by Bethe ansatz \cite{bethe1931zp};
the ground state is an insulator and there is no phase transition for $U>0$ \cite{lieb1968prl}.
Namely, the system preserves the full symmetry over the full range of $U>0$.
We will see how the choice of the one-body potential affects the physical quantities,
especially the ES.
\par
Here we consider $N=120$ sites and $N_{\mathrm{A}}=4$ impurity sites,
which are located at the center of the system.
We provide the initial value of $\bm{u}_{i}$ for the DET-Full and DMET-Full as $\bm{u}_{i}=5\times10^{-3}(0,0,(-1)^{i})$,
which favors an antiferromagnetic spin configuration.
Other elements are initially set to zero.
Figure \ref{fig2}(a) shows the ground-state energy per site for $N=120$ and $N_{\mathrm{A}}=4$,
together with the DMRG data for $N=40$ and $m=800$ under the antiperiodic boundary condition,
where $m$ is the maximum number of a bond dimension.
One can see that all the results are in good agreement with the DMRG energy.
The difference between the DMET (DET) and DMRG is at most $5\times10^{-3}$ as shown in the bottom panel of Fig. \ref{fig2}(a).
We also calculate the double occupancy defined as
%%%%%%%%%%%%%%%
\begin{equation}
n_{d}
=
\frac{1}{N_{\mathrm{A}}}
\sum_{i=1}^{N_{\mathrm{A}}}
\braket{\hat{n}_{i,\uparrow}\hat{n}_{i,\downarrow}}_{\mathrm{imp}}
.
\end{equation}
%%%%%%%%%%%%%%%
The increase of on-site correlation suppresses $n_{d}$ \cite{parcollet2004prl}.
In the exact solution, the double occupancy is related to the ground-state energy as $n_{d}=(\partial E/\partial U)/N$.
Figure \ref{fig2}(b) shows the double occupancy together with the DMRG data.
All types except the DET-SU(2) reproduce the DMRG result with high accuracy. 
It is found that in the case of the DET-SU(2), 
the optimized one-body potential becomes exactly zero over the full range of $U$, 
which does not happen for other cases. 
The symmetry-breaking of the one-body potential gives higher accuracy for the above two quantities. 
%%%%%%%%%%%%%%%%%%%%%%%%%%%%%%%%%%%%%%%%%%%%%%%%%%%%%%%%%%%%%%%%%%%%%%%%%%%%%%%%%%%%%%%%%%%%%%%%%%%%%%%%%%%%%%%%%%%%%%%%
\begin{figure}[t]
\includegraphics[width=85mm]{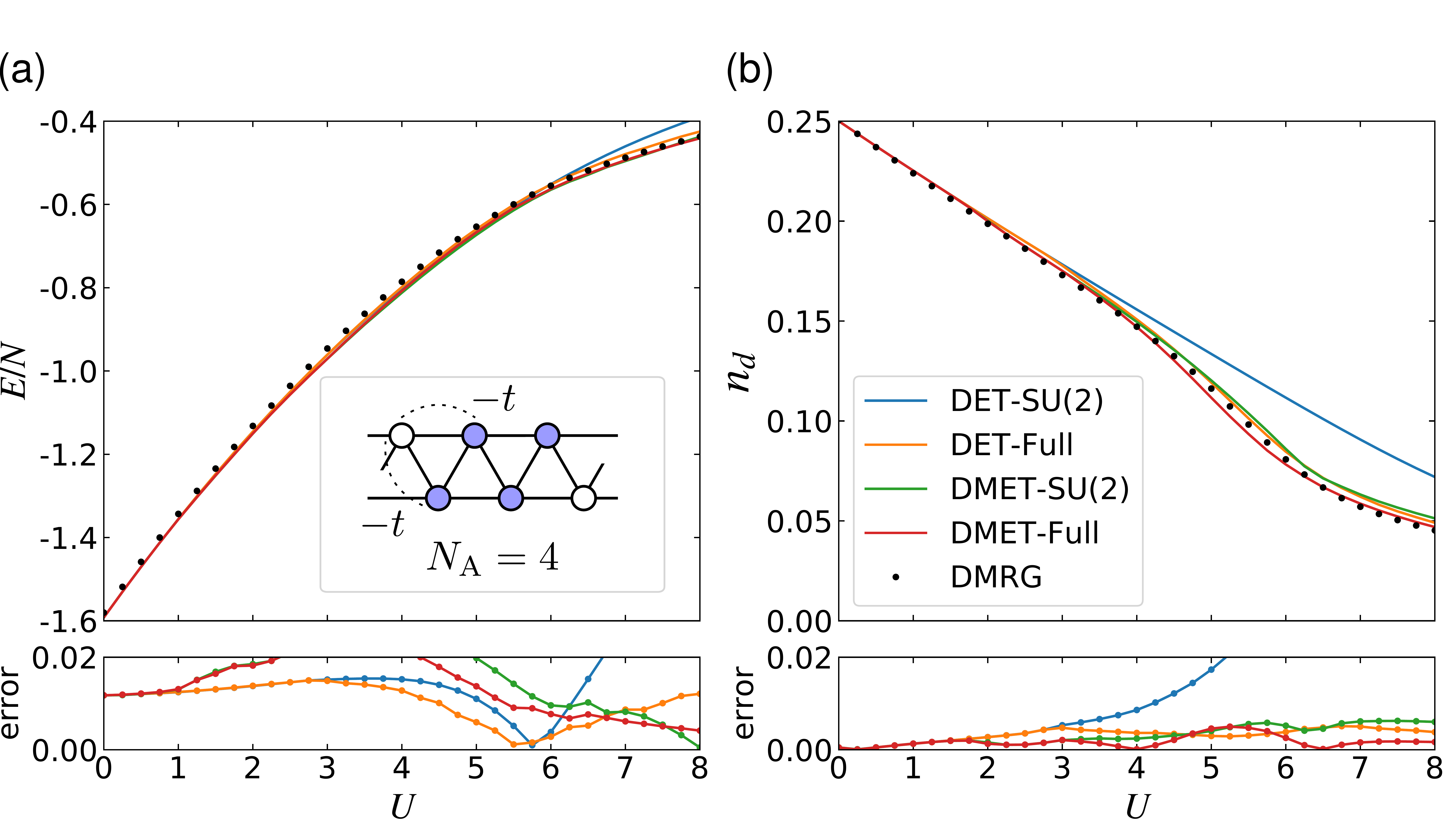}
\caption{(a) Ground-state energy per site and (b) double occupancy of the zig-zag Hubbard chain at half-filling 
as the function of $U$ for $N=120$ and $N_{\mathrm{A}}=4$.
The bottom panel shows the error with respect to the DMRG results for $N=120$ and $m=1000$ under the open boundary condition.
For a large $U$,
the DET-SU(2) result significantly deviates from the DMRG results, while the other types reproduce them.}
\label{fig4}
\end{figure}
%%%%%%%%%%%%%%%%%%%%%%%%%%%%%%%%%%%%%%%%%%%%%%%%%%%%%%%%%%%%%%%%%%%%%%%%%%%%%%%%%%%%%%%%%%%%%%%%%%%%%%%%%%%%%%%%%%%%%%%%
%%%%%%%%%%%%%%%%%%%%%%%%%%%%%%%%%%%%%%%%%%%%%%%%%%%%%%%%%%%%%%%%%%%%%%%%%%%%%%%%%%%%%%%%%%%%%%%%%%%%%%%%%%%%%%%%%%%%%%%%
\begin{figure}[t]
\includegraphics[width=85mm]{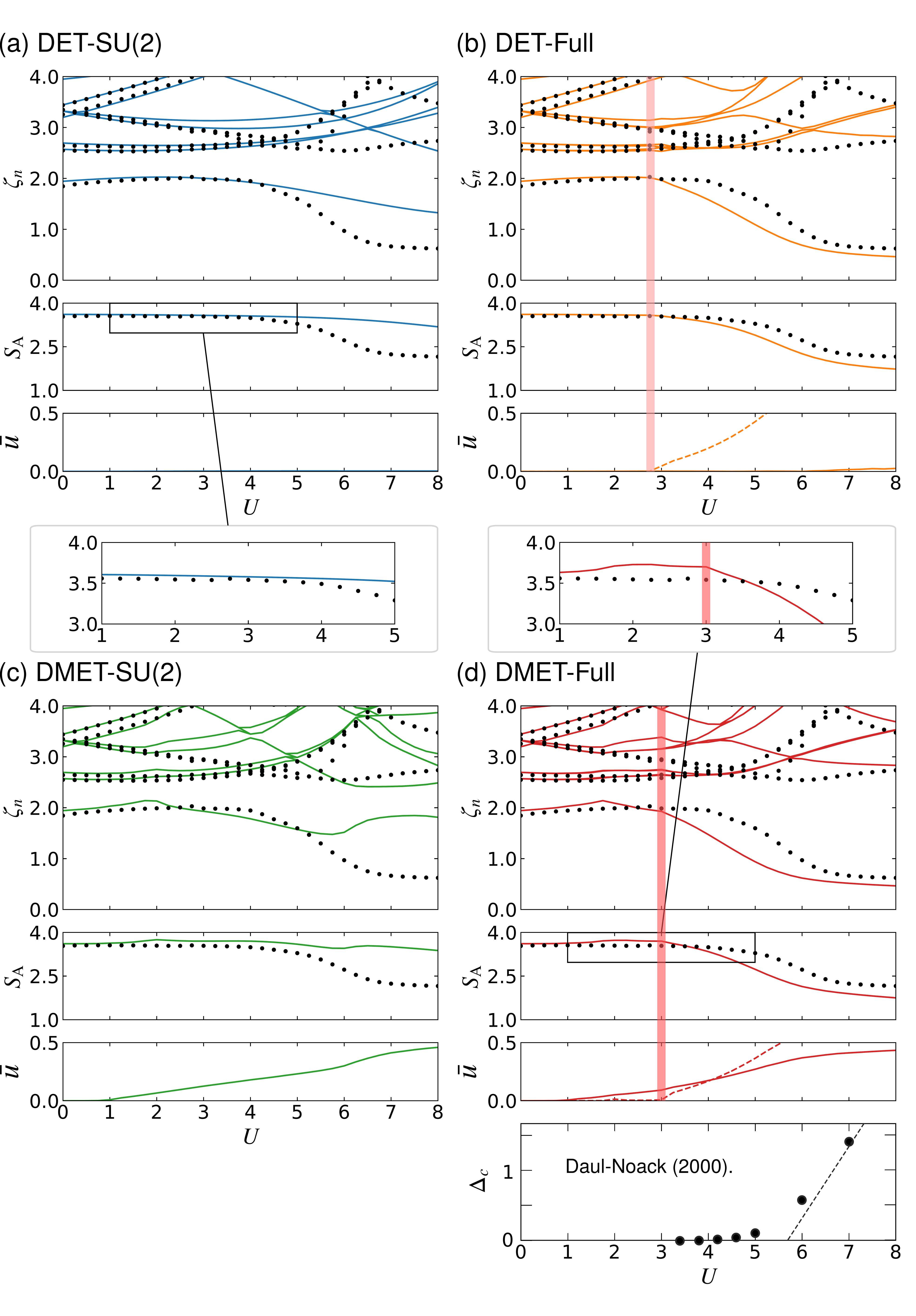}
\caption{ES, EE and the average of the one-body potential of the zigzag Hubbard chain at half-filling as the function of $U$
for (a) DET-SU(2), (b) DET-Full, (c) DMET-SU(2), and (d) DMET-Full.
Data points are the DMRG results.
The DET-SU(2) reproduces the low-energy ES and EE of DMRG for the low-$U$ region,
while in the high-$U$ region, the ES starts to deviate from the data points.
In other types the ES also deviates from the DMRG results.
The phase transition is observed as the onset of the $\bar{u}^{\mu}$ ($\mu=x,y,z$) in (b) DET-Full and (d) DMET-Full.
This transition is also observed as the subtle anomaly of the EE in the inset of (d),
which is not observed in the DET-SU(2) results as shown in the inset of (a).
The charge gap $\Delta_{c}$ in the bottom panel of (d) is extracted from Ref. [\onlinecite{daul2000prb}].}
\label{fig5}
\end{figure}
%%%%%%%%%%%%%%%%%%%%%%%%%%%%%%%%%%%%%%%%%%%%%%%%%%%%%%%%%%%%%%%%%%%%%%%%%%%%%%%%%%%%%%%%%%%%%%%%%%%%%%%%%%%%%%%%%%%%%%%%
\par
However, the entanglement properties behave contrarily. 
Figure \ref{fig3} shows the $U$-dependent ES, EE, and the averaged values of the optimized one-body potential, 
which is defined as
%%%%%%%%%%%%%%%
\begin{equation}
\bar{u}^{\mu}
=
\frac{1}{\dim u^{\mu}}
\sum_{i,j=1}^{N_{\mathrm{A}}}
|u_{i,j}^{\mu}|
,
\label{aveu}
\end{equation}
%%%%%%%%%%%%%%%
where $\dim u^{\mu}$ is the number of elements in $u_{i,j}^{\mu}$ ($\mu=0,x,y,z$).
In Fig. \ref{fig3}(b)-\ref{fig3}(d),
one finds that the ES and EE no longer shows quantitative agreement 
with the DMRG results at $U\gtrsim 0.9$,
where the one-body potential takes finite values (shaded region).
The breaking of the SU(2) spin-rotational symmetry of the potential lifts the degeneracy in the ES,
which can be seen near $U=1.5$ in Fig. \ref{fig3}(b) and \ref{fig3}(d). 
In contrast, for the DET-SU(2) the overall structures of the low-level ES and EE are in good agreement with the DMRG results.
In the DET-SU(2) the one-body potential is exactly zero (see Fig. \ref{fig3}(a)),
namely the trial wave function has the same symmetries with the true ground state.
The other types of potentials break either the SU(2) spin-rotational symmetry, translational symmetry, or both of them.
This symmetry breaking starts to occur when the infinitesimally small charge gap at $U=+0$ starts to
develop rapidly at around $U\gtrsim 0.9$ (see the bottom panel of Fig. \ref{fig3}(d)).
In these regions, the basis taken in by the symmetry breaking potentials has counterparts that together recover the symmetry,
while $E$ and $n_d$ are more accurately evaluated by taking only the symmetry broken part.
This is similar to the situation where the open boundary calculation that breaks the translational symmetry can
describe the ground state with a smaller number of basis than the periodic boundary ones in a DMRG calculation.
Here, one can conclude that for the ground state that continues from $U=0$ and does not break any symmetry,
the DET with the symmetric potential reproduces well the exact ES,
while the other potentials can describe well the extensive physical quantities even though the ES does not match.

%%%%%%%%%%%%%%% Zigzag chain %%%%%%%%%%%%%%%
\subsection{Zigzag chain}
\label{subsec:zigzag}
Let us now consider the case of the half-filled zigzag chain. 
Previous studies suggest that 
there is a metal-insulator transition associated with a dimerization at $U=U_c$ 
\cite{fabrizio1996prb,kuroki1997jpsj,daul1998prb,arita1998prb,daul2000prb,aebischer2001prl,louis2001prb,hamacher2002prl,torio2003prb,gros2005epl,capello2005prl}. 
The value of $U_c$ estimated by the finite scaling analysis on the DMRG calculation is $U_c=3.2$ \cite{daul1998prb}, 
and the one by the variational Monte Carlo method is $U_c=6$ \cite{capello2005prl}. 
This discrepancy is possibly because after the opening of the charge gap in DMRG, 
its amplitude develops very slowly and becomes visible at around $U=5-6$ \cite{capello2005prl}. 
In the strong coupling limit, 
the Hubbard model on the zigzag chain is reduced to the $J_{1}$-$J_{2}$ Heisenberg model with $J_{1}=J_{2}=4t^{2}/U$, 
whose ground state is a singlet dimer that accompanies the spontaneous lattice symmetry breaking
\cite{white1996prb,itoi2001prb,nersesyan1998prl,chitra1995prb,tonegawa1987jpsj,bursill1995jpcm}.
Here we focus on the effect of this symmetry breaking on the physical quantities. 
\par
We consider $N=120$ sites and $N_{\mathrm{A}}=4$ impurities located at the center of the system, and
set the initial value of $\bm{u}_{i}$ for the DET-Full and DMET-Full as $\bm{u}_{i}=5\times10^{-3}(\cos(i\pi/2),\sin(i\pi/2),0)$
referring to the Hartree-Fock calculation in Ref. [\onlinecite{daul2000prb}].
Other initial elements are set to zero.
In Fig \ref{fig4},
we plot the ground-state energy per site and double occupancy as functions of $U$.
The DMRG data shown together is obtained for $N=120$ and $m=1000$ under an open boundary condition,
which favors one of the symmetry broken ground states.
At around $U\gtrsim3.0$,
the ground-state energy and the double occupancy obtained by the DET-SU(2) deviate from the DMRG data,
where the metal-insulator transition occurs and a charge gap begins to develop as well \cite{daul2000prb}.
For other types of potentials, both the exact ground-state energy and the double occupancy are well reproduced.
These results imply that a finite one-body potential is essential to describe
these quantities in the presence of the spontaneous symmetry breaking.
\par
We now turn to the entanglement properties.
Figure \ref{fig5} shows the ES, EE, and the averaged value of the one-body potential. 
The DMRG data points are shown together for comparison. 
We find that the DET-SU(2) and DET-Full with zero potentials well reproduce the ES up to $U\sim3$. 
At around this point, the mean value of the potential $\bar{u}^{\mu}$ (see Eq. (\ref{aveu})) of the DET-Full and DMET-Full
becomes finite, which is seen in a subtle anomaly of the ES.
The off-diagonal elements of the potential start to develop at lower $U$ for the DMET-SU(2) as well,
which modifies the kinetic term in the reference system in a way to generate a spontaneous dimerization.
However, the opening of the charge gap is subtle,
and can only be accurately detected by the onset of $\bar{u}^{\mu}$ in the present framework,
which is consistent with the drop in $n_d$.
In further increasing $U$, however, ES and EE in all types deviate from the DMRG results.
\par
These results imply again that the ES of the DET-SU(2) and DET-Full is reliable in the metallic regime continuing
from the noninteracting point. 
The symmetry breaking phase transition is detected by the DMET-Full and DET-Full
through the change in the structure of the one-body potential,
which is not restricted by symmetries.
%%%%%%%%%%%%%%%%%%%%%%%%%%%%%%%%%%%%%%%%%%%%%%%%%%%%%%%%%%%%%%%%%%%%%%%%%%%%%%%%%%%%%%%%%%%%%%%%%%%%%%%%%%%%%%%%%%%%%%%%
\begin{figure*}[t]
\includegraphics[width=170mm]{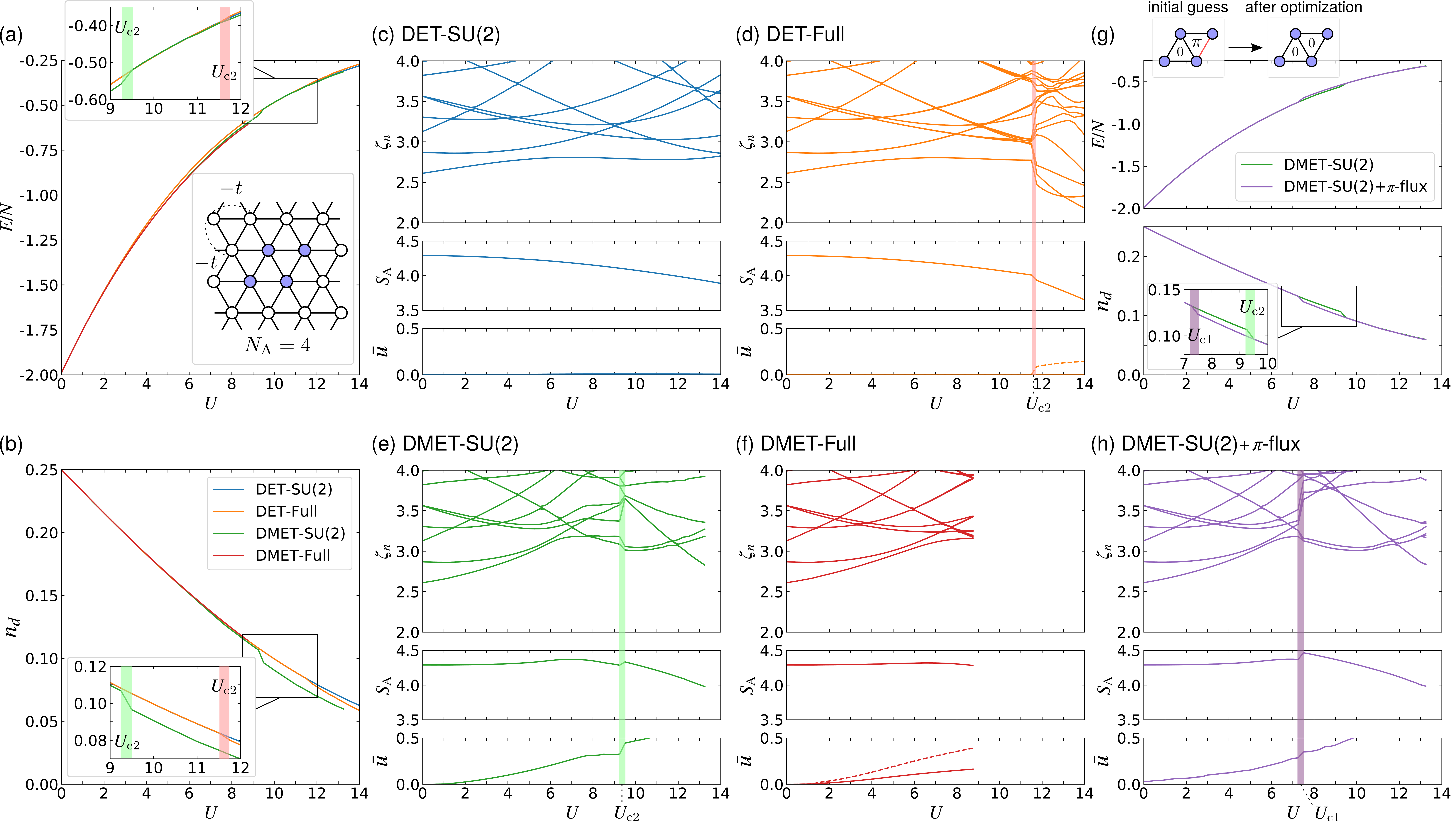}
\caption{Results of the triangular lattice Hubbard model at half-filling. 
(a) Ground-state energy per site. 
The inset shows the energy near the first-order phase transition.
(b) Double occupancy.
We observe the first-order phase transition as the drop of the double occupancy,
which is clearly seen in the inset.
(c-f) ES, EE, and average of the one-body potential as the function of $U$.
The structure of the ES change when crossing the phase transition point.
Note that the one-body-potential optimization of the DMET-Full fails for a large $U$.
(g) Ground-state energy per site and double occupancy obtained by the DMET-SU(2) and DMET-SU(2)+$\pi$-flux.
The drop of the double occupancy is observed at the different $U$.
(h) ES, EE, and average of the one-body potential as the function of $U$ for the DMET-SU(2)+$\pi$-flux.}
\label{fig6}
\end{figure*}
%%%%%%%%%%%%%%%%%%%%%%%%%%%%%%%%%%%%%%%%%%%%%%%%%%%%%%%%%%%%%%%%%%%%%%%%%%%%%%%%%%%%%%%%%%%%%%%%%%%%%%%%%%%%%%%%%%%%%%%%

%%%%%%%%%%%%%%% Triangular lattice %%%%%%%%%%%%%%%
\subsection{Triangular lattice}
\label{subsec:triangular-lattice}

We finally examine the Hubbard model on the triangular lattice.
Identifying the nature of its ground state is a long-standing theoretical challenge.
Many numerical methods have been applied; 
there are indications that the nonmagnetic insulating phase is sandwiched between a metallic and $120^{\circ}$ magnetically-ordered phase 
\cite{morita2002jpsj,watanabe2006jpsj,kyung2006prl,koretsune2007jpsj,clay2008prl,watanabe2008prb,sahebsara2008prl,ohashi2008prl,lee2008prb,liebsch2009prb,galanakis2009prb,yoshioka2009prl,yang2010prl,sato2012prb,kokalj2013prl,tocchio2013prb,yamada2014prb,li2014prb,dang2015prb,shirakawa2017prb,szasz2020prx}.
This intermediate phase has been considered as a candidate of the quantum spin liquid, 
whereas its nature, e.g. whether there exists a spin gap or not, 
whether there is a coexisting nonmagnetic chiral order, what kind of spin liquid it should be,
still remains controversial. 
Clarifying this difficult issue is out of the scope of the present paper,
while one can see whether the phase transition is detected within the present scheme. 
The Mott transition point $U_{c1}$ and the magnetic transition point $U_{c2}$ are evaluated in various methods;
the first path integral renormalization group (PIRG) study gives $U_{c1}=5.2\pm 0.2$ showing a jump in the double occupancy \cite{morita2002jpsj}. 
Later PIRG gives $U_{c1}=7.4$ and $U_{c2}=9.2$ \cite{yoshioka2009prl}. 
For the variational cluster approximation (VCA) smaller values, $U_{c1}=6.3$--$6.7$ and $U_{c2}=8$, are 
observed \cite{sahebsara2008prl,yamada2014prb} possibly because of the cluster-dependent character of the method. 
The cylindrical DMRG in Ref. [\onlinecite{shirakawa2017prb}] up to 48 sites keeping the aspect ratio closer to 1
yield $U_{\mathrm{c1}}=7.55$--$8.05$ and $U_{\mathrm{c2}}=9.65$--$10.15$
detected by the discontinuity in $n_{d}$ and entanglement gap, respectively. 
Another cylinder DMRG with a maximum circumference of 6 with an infinitely long leg 
using the matrix-product-state construction has $U_{c1}=8.5$ and $U_{c2}=10.6$. 
From the scaling analysis, it is empirically known that for the small system size, 
keeping the aspect ratio uniform gives more accurate/reliable numerical results \cite{sandvik2012,chisa2013}.
We thus expect these values to fall at around $U_{c1}=7.4$--$7.8$ and $U_{c1}=9$--$10$. 
\par
Let us consider $N=10\times12$ sites and $N_{\mathrm{A}}=2\times2$ impurities at the center of the system.
We assume $\bm{u}_{i}=h(\cos\bm{Q}\cdot\bm{r}_{i},\sin\bm{Q}\cdot\bm{r}_{i},0)$
and set the initial value of $h$ for the DET-Full and DMET-Full as $h=5\times10^{-3}$, which favors the $120^{\circ}$ spin configuration.
Other initial elements are set to zero. 
Figures \ref{fig6}(a) and \ref{fig6}(b) show the ground-state energy and the double occupancy, respectively.
Both quantities show singularities in the DET-Full and DMET-SU(2) results. 
The discontinuity in Fig. \ref{fig6}(b)
indicates the existence of the first-order phase transitions at $U_{\mathrm{c}2}\simeq 9.25$
and $U_{\mathrm{c}2}\simeq 11.75$ for the DMET-SU(2) and DET-Full, respectively. 
Note that the DMET-Full calculation, which can be regarded as the combination of the DMET-SU(2) and DET-Full, does not converge for large $U$.  
\par
Figures \ref{fig6}(c)--\ref{fig6}(f) show the ES, EE, and average of the one-body potential. 
The transition points observed above manifests themselves as 
the discontinuities in the ES and the jump in the EE.
Although we do not have a reference result by other methods, 
the comparison between them give some clue to understand their overall tendency. 
The ones by the DET-SU(2) and DET-Full below $U_{c2}$ obtained in the absence of the one-body potential 
are in good agreement with each other, 
which should mimic the exact ES of the paramagnetic metallic phase. 
The DET-Full detects the direct symmetry breaking from the paramagnetic to the 120$^\circ$ phase 
which can be interpreted as $U_{c2}$, 
while the ES and EE in the symmetry broken 120$^\circ$ phase may no longer be reliable. 
The DMET-SU(2) and DMET-Full do not give a suitable description of the paramagnetic metallic phase,
but can detect instability to the Mott phase at $U_{c2}$. 
Here, although the DMET-SU(2) did not adopt the 120$^\circ$-type potential, we interpret its anomalous point as $U_{c2}$, 
because the ES above this point resembles those of the DMET-SU(2)+$\pi$-flux which we see shortly.
Unfortunately, $U_{c1}$ is missing, since 
the potentials we have adopted is apparently not suitable for the description of the intermediate phase. 
\par
Referencing the mean-field ansatz for the quantum spin liquid, 
we also consider the initial value of the potential in the DMET-SU(2) as the $\pi$-flux state, which we denote as DMET-SU(2)+$\pi$-flux; 
one of the $u_{i,j}$ is set to $2t$ to introduce the $\pi$-flux on a triangle unit. 
Figure \ref{fig6}(g) shows the ground-state energy and double occupancy of the DMET-SU(2) and DMET-SU(2)+$\pi$-flux. 
We find that after the optimization, 
the resulting potential has no $\pi$-flux but is still different from that of the DMET-SU(2). 
The double occupancy jumps at $U_{c1}\sim 7.25$, 
where the singularity of the ES and EE is also observed in Fig. \ref{fig6}(h). 
At $U\gtrsim 10$, the ES of the DMET-SU(2)+$\pi$-flux are in good agreement with the DMET-SU(2).
\par
These results indicate that the existence of the phase transitions and their locations strongly depends on the choice of the potential. 
Although, one cannot determine quantitatively, e.g. by comparing the energy, 
which type of the potential describes better the target phase of matter,
one can conclude the following. 
Even though the ES differs between potentials and optimizations, 
the energy and the double occupancy do not differ much. 
We also notice that the anomalies of ES and EE are observed as some sort of instability to the given types of potentials. 
In fact, $U_{c1}\sim7.25$ obtained by the $\pi$-flux instability in the DMET-SU(2)+$\pi$-flux 
and $U_{c2}\sim9.25$ by the DMET-SU(2) are both in good agreement with the previous reports $U_{c1}=7.4-7.8$ and $U_{c1}=9-10$. 

%%%%%%%%%%%%%%%%%%%%%%%%%%%%%%%%%%%%%%%%%%
%%%%%%%%%%%%%%% conclusion %%%%%%%%%%%%%%%
%%%%%%%%%%%%%%%%%%%%%%%%%%%%%%%%%%%%%%%%%%
\section{conclusion}
\label{sec:conclusion}
To clarify the applicability and limitation of the DMET,
we have applied several variants of the one-body potential and optimization scheme to 
the 1D, zigzag, and triangular lattice Hubbard models at half-filling. 
We tried the potentials not restricted by symmetries and the SU(2) symmetric ones, 
and adopted the DET and DMET with different optimization schemes;
the DET tries to fit only the diagonal elements of the one-body density matrix 
of the reference Hamiltonian and the impurity Hamiltonian, while the DMET considers 
also the off-diagonal elements. 
\par
By comparing the results with those of the DMRG for the 1D and zigzag cases, 
we have shown that the DET-SU(2) which practically yields zero-potentials and adopts the noninteracting basis set 
reproduces well the ES of the phases at $U>0$ that continues from the noninteracting limit. 
The symmetry-breaking transition point with the subtle charge-gap opening can be detected by the 
emergent asymmetries in the optimized one-body potentials. 
For the triangular lattice where the reference solution is lacking, one needs to apply several types of 
potentials and check the instabilities to the states whose features are encoded in the shape of the potentials. 
Such instabilities are detected by the change in the optimized one-body potentials, and accordingly by the 
discontinuities in the ES and EE. 
Although we have particularly focused on the models with frustrated geometry, 
the overall features of DMET/DET do not depend much on these geometries.
\par
To summarize, the DET with the symmetric potential is useful for the description
of the weakly interacting correlated models,
and the systematic trials on the evaluation of ES using different potentials would serve as a marker for phase transition points. 
However, the lack of variational principles still makes it difficult to judge which of the choices would give the better results. 
%%This does not matter much for the conventional Landau's symmetry breaking phases where the energy and the order parameters are enough, whereas
The difficulty arises particularly for the characterization of exotic phases like topologically-ordered phases
where the ES plays an essential role. 
The issue resolves either if one is able to reasonably increase the size of the impurity toward the exact limit $N/2$, 
or by developing an algorithm that legitimates the optimization of the potential, which are not available at present. 
However, the DMET/DET are still the first to detect the ES of strongly correlated electrons handily, 
which gives important clue to understand the most interesting situations in condensed matters.

%%%%%%%%%%%%%%%%%%%%%%%%%%%%%%%%%%%%%%%%%%%%%%%
%%%%%%%%%%%%%%% acknowledgments %%%%%%%%%%%%%%%
%%%%%%%%%%%%%%%%%%%%%%%%%%%%%%%%%%%%%%%%%%%%%%%
\begin{acknowledgments}
The authors thank Garnet Kin-Lic Chan for helpful discussions.
M.K. performed the DMRG calculations using ITensor library \cite{fishman2020arxiv}.
This work was supported by JSPS KAKENHI Grants
No. JP17K05533,
No. JP18H01173,
and No. JP17K05497.
M. K. was supported by Grant-in-Aid for JSPS Research Fellow (Grant No. 19J22468).
\end{acknowledgments}

%%%%%%%%%%%%%%%%%%%%%%%%%%%%%%%%%%%%%%%%%%%%
%%%%%%%%%%%%%%% bibliography %%%%%%%%%%%%%%%
%%%%%%%%%%%%%%%%%%%%%%%%%%%%%%%%%%%%%%%%%%%%
\bibliography{biblio}

\begin{thebibliography}{117}
\bibitem [{foo({\natexlab{a}})}]{footnote1}%
  \BibitemOpen
  \href@noop {} {\bibfield  {journal} {\bibinfo  {journal} {More precisely, the
  reproduction of the exact ES does not necessarily guarantee the exact reduced
  density matrix since it can have different eigenvectors. Such a situation is,
  however, less likely to occur}\ } ({\natexlab{a}})}\BibitemShut {NoStop}%
\end{thebibliography}
\bibliographystyle{apsrev4-1_mk}

\end{document}